# Study of the neutron quantum states in the gravity field

V. V. Nesvizhevsky, A. K. Petukhov, H. G. Börner (Institut Laue-Langevin, Grenoble, France)
T. A. Baranova, A. M. Gagarski, G. A. Petrov (Petersburg Nuclear Physics Institute, Gatchina, Russia)
K. V. Protasov (Laboratoire de Physique Subatomique et de Cosmologie, IN2P3-CNRS-UJF, Grenoble, France)
A. Yu. Voronin (Lebedev Institute, Moscow, Russia)
S. Baeßler (University of Mainz, Germany)
H. Abele (University of Heidelberg, Germany)
A. Westphal (DESY, Hamburg, Germany)
L. Lucovac (Université Joseph-Fourier, Grenoble, France)

**Abstract**

We have studied neutron quantum states in the potential well formed by the earth's gravitational field and a horizontal mirror. The estimated characteristic sizes of the neutron wave functions in the two lowest quantum states correspond to expectations with an experimental accuracy. A position-sensitive neutron detector with an extra-high spatial resolution of ~2 μm was developed and tested for this particular experiment, to be used to measure the spatial density distribution in a standing neutron wave above a mirror for a set of some of the lowest quantum states. The present experiment can be used to set an upper limit for an additional short-range fundamental force. We studied methodological uncertainties as well as the feasibility of improving further the accuracy of this experiment.

## 1. Introduction

The quantum states of a particle with mass $m$ in the earth's gravitational field with acceleration $g$ above an ideal horizontal mirror are described by the Schrödinger equation that is analytically solved in textbooks on quantum mechanics [1-5]. One possibility for measuring such states for neutrons has been discussed in ref. [6]. The lowest quantum state of neutrons in such a system was observed in a recent experiment at the Institut Laue-Langevin [7, 8].

The present work aims to confirm the existence of this phenomenon and to study it in more detail. Both the experimental installation and the method of measurement remain similar to those used in our previous experiment. However, the instrumental resolution has been improved by improving scatterer positioning. Statistical accuracy has also been improved by optimising neutron transport in front of the experimental installation as well as by more efficient use of the beam time thanks to the complete automatisation of this measurement. The methodological uncertainties of this method as well as possibilities for reducing them were investigated.

This article is organized as follows: Section 2.1 describes new features of the experimental installation compared to that used in ref. [8]. Section 2.2 presents the spectral measurement of the horizontal velocity components. The quality of the mirrors used is studied in section 2.3. Section 2.4 considers the factors responsible for the spatial resolution of our method of measuring the neutron quantum states as the neutrons are transmitted through a slit between the mirror and the scatterer. Section 3.1 presents the measurements with different horizontal velocity components and with different scatterers. Section 3.2 deals with the principles of neutron losses in scatterers/absorbers. Agreement between the experimental data and the model is discussed in section 3.3. Section 3.4 deals with the measurement performed with two subsequently installed scatterers – the first to shape the neutron spectrum (it selects 3-4 lower quantum states), the second to analyze the resulting spectral distribution. In section 3.5, we discuss the upper limit for an additional short-range force resulting from this experiment. Finally, section 4 presents the results of a direct measurement of the probability density to observe neutrons as a function of height above mirror. This measurement was carried out using a position-sensitive neutron detector of extra high spatial resolution of ~2 μm.

The precise analytical solution for the corresponding Schrödinger equation contains Airy-functions. A more transparent and simple solution could be obtained in the quasi-classical approximation [1-3], which is valid for the given potential with a high degree of accuracy of ~1% even for the lowest quantum state. Thus, in accordance with the Bohr-Sommerfeld formula, the neutron energy in quantum states $E_n (n=1,2,3...)$ is equal to:

$$E_n \cong \sqrt[3]{\left(\frac{9 \cdot m}{8}\right) \cdot \left(\pi \cdot \hbar \cdot g \cdot \left(n - \frac{1}{4}\right)\right)^2} \; . \qquad (1)$$

The precise expression for the energy values $E_n$ has the same property as eq.(1): it depends only on $m$, $g$, the Planck constant $\hbar = 6.6 \cdot 10^{-16} eV \cdot s$ and the quantum number $n$. On the other hand, a mirror can be approximated to an infinitely high and sharp potential step compared to other characteristic parameters of the problem. Note that the neutron energy in the lowest quantum state (eq.(1)) $\sim 10^{-12}$ eV is much lower than the effective potential of a mirror $\sim 10^{-7}$ eV, and the range of increase of this effective potential $\sim 1$ nm is much shorter than the neutron wavelength in the lowest quantum state $\sim 10$ μm.

In classical mechanics, a neutron with energy $E_n$ in a gravitational field can rise to the maximum height of $z_n = E_n/mg$. In quantum mechanics, the probability of observing a neutron in an $n$-th quantum state with an energy $E_n$ at a height $z$ equals to the square of the modulus of its wave function $|\psi_n(z)|^2$ in this quantum state (the precise wave functions can be found in refs. [1-5] or in refs. [7, 8]). Formally, this value is not equal to zero at any height $z > 0$. However, as soon as a height $z$ exceeds some critical value $z_n$, specific for every $n$-th quantum state and equals the height of the neutron classical turning point, the probability of observing a neutron approaches zero exponentially fast. This purely quantum effect of neutron penetration into a classically not-allowed region is called the tunneling effect.

An asymptotic expression for the neutron wave functions $\psi_n(z)$ at large heights $z > z_n$ [3] is:

$$\psi_n(\xi_n(z)) \to C_n \cdot \xi_n^{-\frac{1}{4}} \cdot Exp[-\frac{2}{3} \cdot \xi_n^{\frac{3}{2}}], \text{ if } \xi_n \to \infty, \qquad (2)$$

where $C_n$ are known normalization constants;

$$\xi_n = \frac{z}{z_0} - \lambda_n; \qquad (3)$$

and $z_0$ is the characteristic scale for the gravitational quantum states, which is equal to

$$z_0 = \sqrt[3]{\frac{\hbar^2}{2 \cdot m^2 \cdot g}}. \qquad (4)$$

For neutrons at the Earth's surface this value is equal to 5.87 μm.
The values $\lambda_n$ (zeros of the Airy-function) define the quantum energies $E_n = mgz_0\lambda_n$. For the 7 lowest quantum states they are equal to:

$$\lambda_n : \{2.34, 4.09, 5.52, 6.79, 7.94, 9.02, 10.04 ...\}. \qquad (5)$$

The corresponding classically allowed heights are:

$$z_n = E_n/mg = \lambda_n \cdot z_0. \qquad (6)$$

As soon as such a height $z_n$ is reached, the neutron wave function $\psi_n(z)$ starts approaching zero exponentially fast. For the 7 lowest quantum states they are equal to:

$$z_n = \{13.7, 24.0, 32.4, 39.9, 46.6, 53.0, 58.9 ...\} \text{ μm}. \qquad (7)$$

Such a wave-function shape has allowed us to define a method for observing neutron quantum states: we measure their transmission through the narrow slit $\Delta z$ created between a horizontal mirror below and a scatterer/absorber above (from here on generally referred to simplify as a scatterer). If a scatterer is much higher than the turning point for the corresponding quantum state $\Delta z \gg z_n$, the neutrons pass through the slit without significant losses. As the slit size decreases the scatterer starts approaching the neutron wave function $\psi_n(z)$ and the probability of neutron loss increases. If the slit size is smaller than the characteristic size of the neutron wave function in the lowest quantum state $z_1$, the slit will not be transparent for neutrons. It was this phenomenon that was measured in our previous experiment [7, 8].

## 2. Method of the measurement

### 2.1. Experimental installation

The experimental installation and the method of measurement were analogous to those used in our previous work [7-9]. We will therefore just note (where appropriate) the new features; a detailed description of the standard measuring procedure and that of the experimental installation can be found in our previous publications.
A schematic view of the experimental setup is presented in Figure 1.
The experiment consists of the measurement of the neutron flux through a slit between a mirror and a scatterer as a function of the slit size. Slit size could be finely adjusted and precisely measured. The neutron flux in front of the experimental installation (in Figure 1 on the left) is uniform over height and isotropic over angle, in the ranges, which exceed slit size and angular acceptance of the spectrometer respectively, by more than one order of magnitude. The spectrum of the neutron horizontal velocity components is shaped by the input collimator with two plates; these could be adjusted independently to a required height. A low-background detector measures the neutron flux at the spectrometer exit. Ideally, the vertical and horizontal neutron motions are independent. This is valid if neutrons are reflected specularly from the horizontal mirror, and if any influence of a scatterer, or of any other force, is negligible on those neutrons penetrating the slit. If this holds, the horizontal motion of the neutrons (with an average velocity of ~5 m/s) is determined by classical laws, while in the vertical direction quantum motion is observed with an effective velocity of a few centimeters per second and with a corresponding energy of a few peV ($10^{-12}$ eV). The degree of validity of each condition is not a priory guaranteed and should be verified in related experiments.
The experimental set-up varied from that used for previous measurements as follows: 1) The accuracy and reliability of the scatterer positioning is considerably improved by the use of a capacitor method to measure the size of the slit between mirror and scatterer (the positioning system itself was not changed); 2) The statistical sensitivity of the new experiment is improved by optimising neutron transport in front of the spectrometer as well as by the complete automatisation of the measurement; 3) A new neutron detector with even lower background was available; 4) New optical elements (mirrors and scatterers) had been built. They will be described below.

### 2.2. Spectral measurement of the neutron horizontal velocity components along the neutron beam axis

The spectrum of neutron horizontal velocity components was measured in the same way as in ref. [8]. This method is based on a considerable parabolic curvature of cold neutron trajectories in the earth's gravitational field. The free path between the input collimator and a mirror was equal to $l = 10.5$ cm. The mirror length along the neutron beam was $L = 12$ cm. The length of the scatterer along the neutron beam was equal to 10 cm. The slit size between a mirror and a scatterer during measurements of the neutron horizontal velocity components was equal to 150 μm. The macroscopic flatness of the scatterers was better than 0.5 μm. The scatterer surfaces were scanned using standard techniques, in particular using an atomic-force microscope. An example of such a measurement at a small area of one scatterer is shown in Figure 2.

The scatterer surfaces were coated with a copper layer of 0.2 μm thickness by magnetron sputtering. This coating provided an upper electrode for the capacitors used to measure the distances between mirrors and scatterers. Average scatterer roughness (defined as a half-width of the roughness distribution at a half-height) was equal to 0.7 μm. This was sufficient to effectively scatter neutrons in non-specular directions and thus to increase the vertical component of their velocity, so that the increased frequency of their collisions with a mirror and a scatterer allowed their rapid loss. Figure 3 illustrates an integral method to measure the neutron horizontal velocity component spectrum along the neutron beam axis. As one can see, the neutrons penetrating the slit between mirror and scatterer move, in the classical approximation, along parabolas which do not rise higher than the scatterer height but could rise at least as high as the mirror height. This condition, as well as the small size of the slit between mirror and scatterer (compared to the size of the slit in the entrance collimator) allows one to relate a neutron horizontal velocity component along the neutron beam axis $V_{hor}$ uniquely to the difference $\Delta l$ between the neutron turning point height and that of the neutron trajectory at the input collimator. The flight time $\tau_l$ between the collimator and mirror is equal to:

$$\tau_l = \frac{l}{V_{hor}} \quad (8)$$

The difference between the neutron trajectory height at the input collimator and the height of the scatterer is equal to:

$$\Delta l = \frac{g \cdot \tau_l^2}{2} \quad (9)$$

Therefore:

$$V_{hor} = l \cdot \sqrt{\frac{g}{2 \cdot \Delta l}} \quad (10)$$

The results of a measurement of the spectrum of neutron horizontal velocity components along the neutron beam axis are shown in Figure 4.

As one can see in Figure 4, the two methods used to measure the spectrum of neutron horizontal velocity components – with the upper plate and with the lower plate – provide analogous results. For smaller velocities, the measurement made with the upper plate is more statistically precise; for higher velocities that with the lower plate is more precise.

**2.3. Study of the neutron specular reflections from mirrors**

The neutron specular reflections from mirrors were studied using several methods. The large-scale waviness of the mirror surfaces was measured using the light diffraction method (if significant it would produce small-angle neutron scattering in the main experiment). The micro-scale roughness of the mirror surfaces was measured using X-ray scattering (if significant it would produce scattering of neutrons to large angles in the main experiment). Both measurements showed that the quality of the mirrors was sufficiently high to rule out a possibility of significant methodical errors in the present experiment at the current accuracy level [10].

However, the most direct and methodologically transparent experiment to study mirror surfaces should use neutrons with wavelengths close to those in the main experiment. As already mentioned, the expected low probability of neutron loss from specular trajectories is almost impossible to measure at a singular neutron reflection from a surface. In order to increase the observable effect of neutron loss from specular trajectories, we transmitted a neutron beam through a small slit between two parallel mirrors at an angle to the mirror surfaces: this allowed neutrons to be reflected many times. A schematic view of this measurement is shown in Figure 5.

If the specular reflection probability were equal to unity, then the neutron flux $F(\Delta z)$ would be proportional to the slit size (neglecting quantum effects in such a system). However, any neutron loss would decrease the penetrating flux, in particular if the slit size is small and neutrons are reflected many times. In a first approximation, the neutron flux is equal to:

$$F(\Delta z) = \alpha \cdot \Delta z \cdot (1 - K_{loss})^{L \cdot \varphi / \Delta z}, \qquad (11)$$

where $K_{loss}$ is the probability of neutron loss from specular trajectories (given the simplification that it does not depend on the incidence angle and on the neutron velocity), $L$ is the mirror length, $\varphi$ is the average incidence angle of the neutron beam relative to a mirror surface, $\alpha$ is the normalization coefficient and $\Delta z$ is the slit size. Or:

$$K_{loss} = 1 - \sqrt[\frac{L \cdot \varphi}{\Delta z}]{\frac{F(\Delta z)}{\alpha \cdot \Delta z}} \qquad (12)$$

The coefficient $\alpha$ could be estimated as $\alpha \cong F(\Delta z)/\Delta z, \ \Delta z \to \infty$.

The results of such a measurement (in accordance with eq. (12)) are shown in Figure 6 for two cases. The bottom mirror is of polished glass $L = 12$ cm, used in the main experiment; the top mirror is an identical mirror coated with a copper layer of 0.2 μm thickness by magnetron scattering (in the first case) or coated with a layer of Ti-Zr-Gd (in the second case). The distance between the mirrors was measured by the capacitance method and adjusted using piezo-elements, in a way analogous to the main experiment (the procedure used to measure distances is described below in more detail). The angle φ between the neutron beam axis and the mirror surface was equal to $2.5 \cdot 10^{-2}$ rad, which provided many subsequent collisions of neutrons with the mirror surfaces.

If the probabilities of neutron loss from specular trajectories $K_{loss}$ were equal for the glass and copper surfaces, then $K_{loss}$ would be equal to $(8\pm1)\cdot 10^{-3}$ per collision (Figure 6). It is natural to assume that the glass surface was more specular than the copper surface, therefore $K_{loss}$ for this surface is even lower. In any case, $K_{loss}$ was lower than $(1.6\pm 0.2)\cdot 10^{-2}$ per collision even assuming ideal specular properties of the mirror with copper coating. The smaller loss probability $K_{loss}$ measured at smaller slit sizes indicates that the average value $K_{loss}$ in eq.(11) cannot be used precisely if the number of neutron collisions with walls is so high that the angular distribution of neutrons in the slit changes. In any case, the resulting probability of neutron loss from specular trajectories was much lower than the reciprocal number of neutron collisions with the mirror in the main experiment (in the quasi-classical approximation).

In an analogous measurement with a glass mirror, coated with an anti-reflecting layer of Ti-Zr-Gd in the proportion 54%-11%-35% and a thickness of 0.2 μm (which is usually used to produce polarizers for cold neutrons [11]) the probability of neutron loss from specular trajectories was equal to $(1.8\pm0.3)\cdot 10^{-2}$ per collision, under the assumption that $K_{loss}$ for the glass and gadolinium surfaces are equal. However, as is clear from a comparison with the previous measurement, neutrons were lost in the second measurement mainly due to their non-complete reflection from the mirror with an anti-reflecting coating. Thus the probability of neutron loss from specular trajectories at their one collision with an anti-reflecting coating was equal to ~3.5%. Let us compare the normal-to-surface component of the neutron velocity in this measurement to the critical velocity of the surface material. The average neutron incidence angle was equal to $\sim 2.5\cdot 10^{-2}$, the average longitudinal neutron velocity component was equal to ~6.5 m/s. Thus, the average normal-to-surface component of the neutron velocity was ~15 cm/s. This means that an effective potential barrier of 0.15 neV would be sufficient to totally reflect such neutrons. This value is smaller by a factor of $\sim 10^3$ than the values of an effective potential for pure materials (Ti, Zr, Gd) forming the anti-reflecting coating. Although their proportions were chosen to reduce the resulting potential barrier, compensation of more than by a factor of $10^2$ for the initial potential barrier is doubtful for technological reasons. This means that a residual small potential barrier was apparently sufficient for neutron reflection from the anti-reflecting coating. Even if the real part of the potential barrier were completely compensated, the neutrons would effectively reflect from its imaginary part; this causes the so-called metallic reflection of neutrons from strong neutron absorbers.

### 2.4. Spatial resolution of the method

Our installation is a precision one-component gravitational neutron spectrometer. Let us consider the factors that define its spatial resolution in measuring mode of scanning the neutron density using a scatterer.

These factors are the following: 1) such fundamental factors as finite penetrability of the gravitational potential barrier separating the classically allowed height from a scatterer height. In other words, this is the finite sharpness of the asymptotic decrease of the Airy functions describing the wave functions of neutrons in quantum states. In addition, the resolution is limited by the finite time of observation of neutrons in the spectrometer; 2) the uncertainties of the model for the neutron-scatterer interaction, in particular those caused by deformations of the neutron wave functions by a scatterer; 3) such instrumental limitations as the scatterer positioning uncertainty or the finite width of the spectrum of the horizontal neutron velocity.

Let us consider each of these limitations in turn.

### 2.4.1. Fundamental limits for the spatial resolution and a model describing the neutron-scatterer interaction

We have developed several models to describe the interaction of neutrons with rough scatterers in our experiment. We will here focus on the model of a gravitational barrier separating the classically allowed region from the scatterer; we will analyse other models in further publications. The limits inherent in the present model are discussed at the end of this section.

The most important limit for the spatial resolution of our spectrometer in its neutron density scanning mode with a scatterer follows from the finite sharpness of the gravitational barrier penetrability as a function of scatterer height. Let's consider this phenomenon, using, for instance, a standard approach in nuclear physics as in the theory of nuclear $\alpha$-decay. As we know, the lifetime of such a nucleus is equal to the penetrability of the nuclear potential barrier for an $\alpha$-particle multiplied by the frequency of collisions of an $\alpha$-particle with this barrier. In order to estimate the neutron lifetime in our case, let's assume that the loss of neutrons per time unit is proportional to the probability of their observation at the scatterer height.

The non-disturbed probability density $|\psi_n(\xi(z))|^2$ to observe neutrons at a height $z$ in the classically prohibited region $z > z_n$ is equal to the square of the modulus of the neutron wave function given by eq.(2):

$$|\psi_n(\xi_n)|^2 \to C_n^2 \cdot \xi_n^{-\frac{1}{2}} \cdot Exp[-\frac{4}{3} \cdot \xi_n^{\frac{3}{2}}], \text{ if } \xi_n \to \infty \qquad (13)$$

Formally, the neutron wave function inside a scatterer is different from the non-disturbed wave function given by eq.(13), given the interaction of the neutrons with the scatterer. I we neglect this effect and a factor smoothly depending on $\xi$, we can estimate the probability $P_n^{tunnel}(\Delta z)$ of observing neutrons in $n$-th quantum state inside a scatterer at a height $\Delta z > z_n$ as:

$$P_n^{tunnel}(\Delta z) = \int_{\Delta z}^{\infty} |\psi_n(z)|^2 \cdot dz \approx Exp\left(-\frac{4}{3} \cdot \xi_n^{\frac{3}{2}}\right), \; \xi_n = \frac{\Delta z - z_n}{z_0}, \; \Delta z > z_n \qquad (14)$$

Let's note that the very same expression could be obtained when calculating the probability of neutrons tunneling through the gravitational barrier in the quasi-classical approximation. As can be seen from eq.(14), the probability $P_n^{tunnel}(\Delta z)$ decreases sharply at any scatterer height $\Delta z$ larger than $z_n$. If scatterer height is $\Delta z \leq z_n$, then the probability $P_n^{tunnel}(\Delta z)$ is high and could be taken to be equal to unity:

$$P_n^{tunnel}(\Delta z) = 1, \; \Delta z \leq z_n \qquad (15)$$

Due to the tunneling of neutrons through the gravitational barrier into a scatterer, even if a scatterer height is large, the neutron states are not precisely stationary. The rate of the states' decay, or the reciprocal lifetime of neutrons in $n$-th quantum state $(\tau_n^{abs}(\Delta z))^{-1}$, can be estimated as the frequency of neutron collisions with the gravitational barrier $w_n$ multiplied by the probability of tunneling of neutrons inside a scatterer (eqs.(14),(15)):

$$\frac{1}{\tau_n^{abs}(\xi_n(\Delta z))} \approx \begin{cases} w_n \cdot Exp\left(-\frac{4}{3} \cdot \xi_n^{\frac{3}{2}}\right), \xi_n > 0 \\ w_n, \xi_n \leq 0 \end{cases} \quad (16)$$

The frequency $w_n$ for neutrons with an energy $E_n$ is defined as:

$$w_n = \frac{E_{n+1} - E_n}{\hbar} \approx \sqrt[3]{\frac{m \cdot \pi^2 \cdot g^2}{3 \cdot \hbar \cdot \left(n - \frac{1}{4}\right)}} \quad (17)$$

Let's consider the transmission of neutrons with a horizontal velocity component along the neutron beam axis $V_{hor}$ through a slit between a mirror and a scatterer with a slit size of $\Delta z$. During the passage time through the slit with a length $L$, which is equal to

$$\tau^{pass}(V_{hor}) = \frac{L}{V_{hor}}, \quad (17)$$

the $n$-th neutron state partially decays. A probability of neutron transmission $P_n(\Delta z, V_{hor})$ is equal to:

$$P_n(\Delta z, V_{hor}) = Exp[-\frac{\tau^{pass}(V_{hor})}{\tau_n^{abs}(\Delta z)}]. \quad (18)$$

In our experiment, if the scatterer height is small $\Delta z \leq z_n$ then the lifetime for n-th quantum state $\tau^{abs}(\Delta z)$ (eqs.(16),(17)) is much shorter than the neutron passage time under a scatterer $\tau^{pass}(\Delta z)$ (eq.(18)). Therefore the probability of transmission of neutrons through the slit $P_n(\Delta z, V_{hor})$ (eq.(19)) in such a case is low, and the precise dependence of the tunneling probability (eqs.(14),(15)) on scatterer height $\Delta z$ and on dependence $\tau^{abs}(\Delta z)$ (eqs.(16),(17)) is not of importance. This fact justifies our choice of eq.(15). Finally, the transmitted neutron flux for several quantum states is equal to:

$$F(\Delta z, V_{hor}) = \sum_n F_n(\Delta z, V_{hor}) = F_0 \cdot \sum_n \left( \beta_n \cdot Exp\left(-\frac{L}{V_{hor}} \cdot \sqrt[3]{\frac{m \cdot \pi^2 \cdot g^2}{3 \cdot \hbar \cdot \left(n - \frac{1}{4}\right)}} \cdot \begin{cases} Exp\left(-\frac{4}{3} \cdot \left(\frac{\Delta z - z_n}{z_0}\right)^{\frac{3}{2}}\right), \Delta z > z_n \\ 1, \Delta z \leq z_n \end{cases} \right) \right)$$
(20)

where $\beta_n$ is a population of $n$-th quantum state, and $F_0$ is the normalization coefficient. The sharpness of the dependence $F_n(\Delta z, V_{hor})$ (eq.(20)) on $\Delta z$ in the range $\Delta z > z_n$ defines the accuracy with which it is possible to identify the different quantum states of neutrons in the measurement of the total neutron flux; it therefore defines the best possible spatial resolution of our spectrometer as well.

Evidently, the ideal resolution would be achieved if dependence $P_n(\Delta z)$ resembled a step-function:

$$P_n^{ideal}(\Delta z) = \begin{cases} 1, \Delta z > z_n \\ 0, \Delta z \leq z_n \end{cases} \qquad (21)$$

The "smoothness" of an actual function $P_n(\Delta z)$ is defined, as is seen from eq.(20), by the penetrability of the gravitational barrier.

As one can see in Figure 7, the higher the value $(\Delta z - z_n)$, the sharper the dependence of the quantum level lifetime on $\Delta z$. A "sharper" measuring mode would correspond to longer observation times for neutrons in the slit between mirror and scatterer, i.e. to a larger mirror-scatterer pair or to a smaller horizontal velocity component. However, the possibility of achieving a decrease in neutron velocity or significant increase in mirror-scatterer length is rather limited in practice. This means that the spectrometer resolution in the present experiment is limited by a fundamental physical phenomenon: the quantum tunneling of neutrons through a gravitational potential barrier.

The spatial resolution of our spectrometer could be noticeably improved if only the neutron storage times in quantum states were increased by a few orders of magnitude in a "storage" measuring mode as discussed in refs. [7, 8]. The other methods to considerably increase the spectrometer resolution are based on precision position-sensitive neutron detectors or on a measurement of the frequency of resonance transitions between the quantum states.

The model dependence (20) is obtained for the lowest quantum states neglecting any deformation of the neutron quantum wave functions by a scatterer. In fact a scatterer "pushes out" the wave functions as is common in quantum-mechanical problems with strong absorption. For this reason the "experimental" values $z_n$ obtained with eq.(20) are systematically slightly lower than the expected ones (7). Besides, we used an approximate asymptotic expression for the neutron wave functions. For small values $\xi_n < 1$ this expression should be improved. Nevertheless eq.(20) reflects the most important physical phenomena and could be used to approximate the experimental data:

$$F(\Delta z, V_{hor}) = \sum_n F_n(\Delta z, V_{hor}) = F_0 \cdot \sum_n \left( \beta_n \cdot Exp\left( -\alpha \cdot \frac{L}{V_{hor}} \cdot \begin{cases} Exp\left( -\frac{4}{3} \cdot \left( \frac{\Delta z - z_n}{z_0} \right)^{\frac{3}{2}} \right), \Delta z > z_n \\ 1, \Delta z \leq z_n \end{cases} \right) \right),$$
(22)

where $\alpha$ is the coefficient responsible for a finite scatterer efficiency and for the approximation used in eq.(14). Such a parametrisation describes experimental data with a small number of free parameters. The effective height of a rough scatterer is defined as the average height over a scatterer surface, if the roughness is significantly smaller than $z_0$. If it is higher or comparable to $z_0$ the definition of an actual effective height of a rough scatterer should be defined more rigorously in the framework of a more precise model. We estimate that in the present experiment the total model uncertainty could produce a systematic shift in the values of the turning points of about ±1.5 μm. A similar expression to eq.(22) has been derived in ref. [10].

It is interesting to estimate a minimal uncertainty value in the neutron quantum states' energy in our experiment, which would follow from the uncertainty principle. In order to do that we should compare the observation time for neutrons in our set-up $\tau^{pass}(V_{hor})$ (eq.(18)) with a characteristic quantum-mechanical interval $\Delta\tau_{QM}$, which is equal to the ratio of the Planck constant to the neutron energy in the lowest quantum state:

$$\Delta\tau_{QM} = \frac{\hbar}{E_1} \cong 0.47 ms \qquad (23)$$

For the average horizontal velocity component 6.5 m/s (see Figure 3) this ratio is equal to $\Delta\tau_{QM}/\tau^{pass}(6.5 m/s) \cong 3\%$, which means that the observation time in the performed experiment was sufficiently long for reduce the corresponding uncertainty to below the current accuracy level.

### 2.4.2. Uncertainty in scatterer positioning

The spatial resolution of the method is determined mainly by fundamental reasons, as described above, if the scatterer positioning uncertainty is lower than ~3 µm, which corresponds to the estimated sharpness of the gravitational barrier penetrability as a function of scatterer height.

The distance between the mirror and scatterer $\Delta z$ was measured using a "capacitor" method. A few metallic (aluminum) electrodes (each electrode with a size of 2.0 cm by 1.5 cm and a thickness of 0.2 µm) were applied via thermal evaporation onto the glass flat surface of a bottom mirror. The capacitors formed between these electrodes and the large metallic coating (thickness 0.2 µm) on the lower face of the scatterer were connected in $RC$ electric chains, which defined oscillation frequencies for the relaxation $RC$-generators. Any oscillation frequency was related to a well-reproducible distance between the surfaces via a dependence, which could be rather precisely approximated by a second-power polynomial in the range of interest 5 µm< $\Delta z$ <1 mm. Smaller slits are not transparent for neutrons anyway, nor could they be installed in an automatic mode because of unavoidable dust particles between the mirror and a scatterer. At distances of >1 mm (not relevant in the present experiment) the capacitance between the electrodes is small and the sensitivity of the capacitor sensors decreases.

The unique relation between the oscillation frequency of the $RC$-generator (which includes a corresponding capacity) and the distance between electrodes provides a high reproducibility of the distance measurement (<<1 µm). The complementary procedure of absolute calibration was carried out in three independent ways: 1) Using tungsten wires-spacers of known diameter placed between the mirror and the scatterer; 3) Using a long-focus microscope, which provides an optical image of a slit; 3) Using a precision mechanical device (comparator) allowing the relative translation of a point at the upper surface of a scatterer to be measured.

The results of the calibration measurements are shown in Figure 8. None of the methods used provide contradictory results. However, their accuracy and methodical errors are different: 1) Tungsten wires-spacers provide the most precise and reliable results with reproducibility of better than 0.3 µm for different measurements and/or pieces of wires of equal thickness. The wires were not noticeably deformed by the measurement. Dust particles between a mirror and a scatterer did not produce any additional uncertainty, because the probability of finding a dust particle just under (or above) a wire is low; such an event would be easily identified by repeating the measurement several times and checking its reproducibility. 2) A distance measurement using a long-focus microscope also provides an absolute distance calibration because if fixes the moment when a scatterer touches a mirror. However, the scattering of measured values in such a way was significantly higher than that in the previous case, due in particular to mechanical problems in the measurements. As one can see in Figure 8, both results coincide within the experimental accuracy. 3) A measurement with a mechanical comparator provides a relative accuracy of ~1 μm. Absolute calibration in such a way is not reliable, however, because the contact between the mirror and scatterer surfaces could be imitated by dust particles between the mirror and scatterer, and these can not be avoided completely because of their large square areas.

Application of the most successful method allows us to calibrate an absolute height with an accuracy of ~0.5 μm. However, due to the difference in the square areas of the electrodes and/or different parasitic capacitances, the capacitor sensors produce different frequencies at equal scatterer heights within a range of 5-7 %. For simplicity's sake, we assumed the sensors to be equivalent, an assumption leading to deviations in the precise parallelism of mirror and scatterer. The corresponding uncertainty in $\Delta z$ was equal to ±1.0 μm at $\Delta z \sim 15$ μm and ±1.6 μm at $\Delta z \sim 25$ μm; this was acceptable in the present experiment. However, it could be significantly reduced in future.

### 3. Experimental results

### 3.1. Measurement of the neutron flux through a slit between a mirror and a scatterer as a function of slit height and the horizontal neutron velocity component

The dependence of the neutron flux on the size of the slit between mirror and scatterer as a function of slit size was measured using techniques similar to those in refs. [7, 8]; the optical elements could be more accurately positioned and higher statistical accuracy was possible. We also studied in more detail the dependence of the neutron flux $F(\Delta z, V_{hor})$ (eq.(22)) on the value of a horizontal neutron velocity component and on the type of absorber/scatterer. The dependence of scatterer efficiency on the roughness of the scatterer surface has not yet been investigated. In the present experiment the average roughness of the scatterer surface was equal to 0.7 μm – i.e. a factor of 1.5 smaller than in refs. [7, 8]. Apparently smaller roughness amplitudes correspond to lower scatterer efficiency and lower spectrometer resolution. The quantitative influence of this factor is not yet completely clear and requires further study.

The results of a measurement made with a copper coated scatterer and with both a broad neutron spectrum and the soft fraction of this spectrum are shown in Figure 9. The spatial resolution of the measurement could be improved by softening and better monochromatising the spectrum of the horizontal velocity components.

Figure 10 illustrates a comparison of analogous measurements with different average horizontal neutron velocity components. As one could see in Figure 10, the harder the spectrum of horizontal neutron velocity components along the neutron beam axis, the smaller the value of $X$; where $X$ is the free parameter in the simplified quantum-mechanical dependence, which assumes a presence of the lowest quantum state alone and classical asymptotics at large slit sizes, or, in other words, which approximates the data at small slit heights using the function $F(\Delta z) \sim (\Delta z - X)^{\frac{3}{2}}$. The three curves in Figure 10 approximate the experimental data at small slit sizes, so that the values of $X$ at zero height are equal to: 13.1±0.2 μm for $\overline{V_{hor}}$=4.9±0.2 m/s; 12.20±0.15 μm for $\overline{V_{hor}}$=6.5±0.2 m/s; and 11.90±0.25 μm for $\overline{V_{hor}}$=7.8±0.3 m/s. These values correspond to $\frac{\partial X}{\partial (\overline{V_{hor}})}$=0.4±0.1 μm/(m/s) at $\overline{V_{hor}}$=6.5 m/s, and $\frac{\partial X}{\partial (\tau^{pass}(\overline{V_{hor}}))}$=0.16±0.04 μm/(ms) at $\tau^{pass}$=15 ms.

It is interesting to compare this measured value with its theoretical expectation. In order to do so one should estimate the sharpness of the dependence $F_1(\Delta z, V_{hor})$ on its parameter $V_{hor}$, i.e. $\frac{\partial (F_1(\Delta z, V_{hor}))}{\partial (V_{hor})}$. An analysis of eq.(22) shows that the result is not particularly sensitive to the problem parameters; it is equal to $\frac{\partial X}{\partial (\tau^{pass})}$=0.2 μm/(ms). The theoretical calculation is therefore in reasonable agreement with the experimental result presented above.

### 3.2. The principles of neutron scattering/absorption

In order to improve the spatial resolution of a spectrometer (with a given scatterer-mirror size) or to decrease mirror-scatterer size (with a given spatial spectrometer resolution) the factors defining scatterer efficiency need to be studied. Before the experiment we were considering two methodical approaches: (a) an absorber with large cross-section of absorption or inelastic neutron scattering, as proposed in ref. [6]; and (b) a scatterer, which reflects neutrons away from their specular trajectories thanks to its elastic (but diffusive) scattering properties, as proposed in ref. [12].

The advantage of the absorber option is the simplicity with which one can define the absorber height corresponding to the height of its flat surface. However its efficiency is low: 1) even an ideal absorber with zero effective potential would reflect neutrons of the smallest energy because of the imaginary part of the reflecting potential; 2) because of combination of materials (with typical effective potentials of ~$10^{-7}$ eV) it is practically impossible to compensate for the effective potential with sufficient precision to achieve a potential of ~$10^{-12}$ eV (defined by the parameters of the quantum states of neutrons in the earth's gravitational field).

The scatterer of method (b) on the other hand (based on elastic diffusive neutron scattering from its macroscopically flat and microscopically rough surface) overcomes the problem of the absorber's low efficiency. It scatters neutrons in non-specular directions with a probability close to unity in the quasi-classical approximation. This is true if the scatterer surface roughness is comparable to the neutron wavelength in the quantum states. On the other hand, this condition raises the problem that the effective scatterer height is defined with an uncertainty comparable to the roughness amplitudes. A more precise definition of the scatterer height could be achieved if a more adequate theory of neutron scattering at rough scatterer surfaces could be developed for cases in which the neutron vertical wavelength is comparable to the roughness size and the influence of the gravitational field is taken into account.

We compared different scatterers/absorbers. Figure 11 presents results of two experiments, in particular: 1) one with a rough copper-coated scatterer; 2) another with an analogous rough scatterer coated with an anti-reflecting layer Ti-Zr-Gd.

Figure 11 shows that the results of measurements with rough scatterers coated with copper and Ti-Zr-Gd layers coincide within experimental uncertainty. This means that the dominant property of a scatterer is its ability to scatter neutrons at its roughness but that the absorbing properties themselves of a scatterer are not of importance. The final result is shown in Figure 12.

### 3.3. Description of the experimental results within the model of neutron tunneling through the gravitational barrier

The curve (22) in Figure 12 approximates the experimental data with $\chi^2=0.9$. In this calculation, the critical heights $\{z_n, n>2\}$ corresponded to eq.(7) for pure quantum states and the two lowest quantum heights $z_1$ and $z_2$ were fitted from the experimental data; the quantum state populations $\{\beta_n, n>1\}$ were assumed to be equal to each other, except for the lowest quantum state population $\beta_1$, which was calculated from the experimental data; the scatterer efficiency $\alpha$ was also calculated from the experimental data.

Comparison of this model, for the infinitely good instrumental resolution, with the experimental data shows that: 1) the lowest quantum state population ($\beta_1 \approx 0.7$) is lower than other populations $\beta_i, i>1$, as was also noticed in refs. [7, 8]; 2) the first and second critical heights $z_1$ and $z_2$ are equal to 11.2 μm and 20.2 μm respectively. The finite accuracy of the $\Delta z$ measurement shifts the values $z_1$ and $z_2$ by +1.0 μm and +1.4 μm respectively (as a pure height resolution imitates weaker scatterer efficiency and smoother flux/height dependencies, i.e. false shift of $z_i$ values). As we estimated above, the systematic uncertainty for $z_1$ and $z_2$ values is due to the absolute distance calibration accuracy (±1.0 μm and ±1.6 μm respectively) and the finite accuracy of the model describing neutron scattering (effective uncertainty ±1.5 μm). The statistical uncertainty of these parameters (±0.7 μm) is lower than the systematical uncertainty. Thus the average values of the two lowest critical heights ($z_1^{\exp}=12.2\pm1.8_{syst}\pm0.7_{stat}$ μm) and ($z_2^{\exp}=21.6\pm2.2_{syst}\pm0.7_{stat}$ μm) do not contradict the expected values in eq.(11): ($z_1^{theor}=13.7$ μm) and ($z_2^{theor}=24.0$ μm) within 25 %. As noticed above, the "experimental" values $z_n$ obtained are slightly too low since the correction for the deformation of the neutron wave-functions is not taken into account explicitly. The expected spatial spectrometer resolution, as seen in Figure 12, agrees with the experimental result. The neutron transmission (eq.(22)) depends smoothly on the scatterer efficiency.

### 3.4. An estimation of neutron storage time in quantum states

The measurement of neutron transmission through a slit between two parallel mirrors allows one to estimate the probability of losing a neutron from the specular trajectory at one bounce; the angular resolution of this method is not high enough however to estimate the probability of transitions between the neighboring quantum states. There is another experiment more sensitive to the transitions between the quantum states; a long mirror and two independent scatterers are installed in such a way that the first scatterer shapes the neutron spectrum and the second scatterer analyses the resulting spectrum. A zone of 9 cm length between the mirrors allows for any evolution of the spectrum. The layout of such an experiment is shown in Figure 13 and the results in the vicinity of the height 42 μm are given in Figures 14.

An analysis of the experimental data presented in Figure 14 shows that 1) the neutron spectrum measured with the second analyzing scatterer does not differ noticeably from that measured with the first shaping scatterer, at the slit sizes $\Delta z < 42$ μm; 2) at the height of approximately $\Delta z \approx 42$ μm one observes that the experimental curve is "broken" because neutrons of higher energy are strongly suppressed by the first scatterer (more precisely, this height is defined by the shape of the Airy function corresponding to the 4$^{th}$ quantum state – see eq.(7)); 3) the total proportion of neutrons detected at the second scatterer height $\Delta z \gg 42$ μm is as high as 35% of the amount of neutrons detected at $\Delta z = 42$ μm.

Let us assume a conservative estimate of the lower limit for storage times of neutrons in quantum states. At the very least, a fraction (or all) of the additional neutrons at $\Delta z > 42$ μm passed under the first scatterer at quantum states $n > 4$ and diffracted at its exit. However, let us attribute the whole count rate at $\Delta z > 42$ μm to the transitions between different quantum states in the 9cm region between the two scatterers. Thus, the upper limit for the probability of neutron transitions from non-disturbed quantum states $n \leq 4$ to non-disturbed quantum states $n > 4$ is at least smaller than ~$10^{-1}$ per one (quasi-classical) collision with a mirror, or ~20 s$^{-1}$; this corresponds to the storage time $> 10^2 \cdot \tau_{QM}$. There is a clearly still need for further experimental and theoretical study to determine the origin of neutrons in the quantum states $n > 4$ and to achieve a reliable estimation of neutron storage times in quantum states. Such an upper estimate for the neutron storage time in low quantum states, however, already indicates the principle feasibility of the so-called "storage" method, based on the long storage of neutrons in a closed trap and their resonance transitions from one quantum state to another using, for instance, the oscillations of a reflecting surface, as in ref. [13], or even using the variation of a gravitational field by oscillating a test mass in a vicinity of the experimental installation. Such precision experiments could be used, for instance, to check the electrical neutrality of neutrons, to test the weak equivalence principle for a quantum particle [14, 15], or to search for hypothetical additional short-range fundamental forces [16, 17] violating the weak equivalence principle.

### 3.5. An upper limit for an additional short-range force

This experiment has shown that, within the experimental accuracy, the neutron wave functions in the quantum states correspond well to their theoretical expectation, which assumes that neutrons are affected only by the Earth's gravitational field and a mirror. If an additional short-range force with sufficient strength were to act between a neutron and a mirror, for instance, as proposed in refs. [18, 19, 23], the neutron wave functions would be deformed by this interaction. The absence of such deformation of the wave functions allows us to assign an upper limit for the intensity of such a hypothetical fundamental interaction in the distance range of 1 nm – 10 μm. This limit is analyzed in ref. [20]. In the range of 1 μm – 10 μm numerical limits have been derived in ref. [22]. The smallest distance considered is defined by the mirror roughness and by the thickness of the layer of surface impurity, and, correspondingly, by the finite sharpness of the increase in the mirror's effective potential. The largest distance considered is defined by the characteristic size of the neutron wave functions in the lowest quantum states.

The experiment [8] was neither designed nor optimized to search for additional short-range forces. A dedicated neutron experiment which is much more sensitive than the constraints in ref. [21, 20] will be discussed in a later article.

**4. Measurement of the neutron spatial density distribution using a position-sensitive detector**

The direct measurement of the spatial density distribution in the neutron standing wave above a mirror is preferable to its study using a scatterer with a variable height. The first method is a differential one, because it allows one to measure simultaneously the probability of observing neutrons at any height. The second method is an integral one, because the probability of observing neutrons at a specific height is determined by the difference between two neutron count rates measured at two close scatterer heights. Evidently, the differential method is much more sensitive and provides a much faster gain in the required statistical accuracy. Besides, a scatterer (used in the integral method) inevitably disturbs the measured quantum states: it deforms the wave functions and shifts the energy levels. The finite accuracy of the corresponding corrections causes systematic uncertainties, which finally limit the accuracy of measurement of the quantum states parameters achievable. For these and other reasons, the use of a position-sensitive detector is preferable for directly measuring the probability of observing neutrons above a mirror. However, no neutron detector existed of the type required for the present experiment, with a position resolution of the order of 1 μm. We were therefore obliged to develop the detector and the method of measurement: a plastic nuclear track detector (CR39) with a uranium coating ($^{235}UF_4$) described in ref. [9]. The point of entry of a neutron into the plastic detector was measured, after its chemical treatment, using an optical microscope allowing the detector surface to be scanned along a distance of about 10 cm in a horizontal direction with an accuracy of ~1 μm along a vertical direction. A detailed description of the measurement and the results will be presented in a forthcoming paper. In the present work we illustrate the capability of this method to measure directly the spatial neutron density distribution with an accuracy of 1-2 μm.

The set-up of this experiment was similar to that shown in Figure 13. The populations of the quantum states were measured in a separate experiment with two scatterers using a measured dependence similar to that in Figure 14. Later, the second scatterer was removed and a position-sensitive detector with uranium coating was installed in the place of the gaseous neutron detector (counter). The goal of this measurement was merely to estimate the spatial resolution of such a detector. It was not optimized for the investigation of quantum states of neutrons above a mirror. In particular, we measured a few quantum states simultaneously, in order to improve the statistics and sharpness of increase of neutron count rate at zero height above the mirror. The results obtained with this detector are shown in Figure 15.

In Figure 15: 1) the neutron wave functions in the quantum states are known [1-5]; 2) the populations of the quantum states calculated here correspond to those measured with the two scatterers method (the data in Figure 14 are analyzed using eq.(22)); 3) the detector spatial resolution is assumed to be infinitely high; 4) the zero height is estimated from the experimental data. As evident from the comparison of the experimental data to the theoretical curve in Figure 15: 1) the measured distribution of neutron density above a mirror corresponds well to its theoretical expectation ($\chi^2$=1.0, excluding 3 data point near the sharp edge); 2) the detector spatial resolution is as high as ~2 μm; 3) Even a relatively small variation in the neutron density of about 10 % (expected for a few quantum states) could be measured in this way. This means that the selection of one-two quantum states (see ref. [12]) would allow us to easily identify them and to measure their parameters.

**Conclusion**

In the present work, we investigated the phenomenon of the quantization of the neutron states in the potential well formed by the earth's gravitational field and a horizontal mirror. The conclusions of our work [7, 8] on the presence of such quantization are confirmed with high statistical accuracy and methodological reliability. We measure the characteristic sizes of the neutron wave functions $z_1$ and $z_2$ in the first and second quantum states. The values measured $z_1^{\exp} = 12.2 \pm 1.8_{syst} \pm 0.7_{stat}$ μm and $z_2^{\exp} = 21.6 \pm 2.2_{syst} \pm 0.7_{stat}$ μm agree with the values expected $z_1^{theor} = 13.7$ μm and $z_2^{theor} = 24.0$ μm, calculated on the assumption that neutrons are affected only by the earth's gravity and a horizontal mirror potential. The accuracy of this experiment is defined by its systematical uncertainty: the uncertainty of the scatterer positioning and uncertainties in the model describing the interaction of neutrons with a rough scatterer. In ref. [20] limits for additional short-range forces in the range from 1 nm to 10 μm were derived from the fact that the quantum states of neutrons in the gravitational field were found to be consistent with the standard quantum-mechanical prediction. An improvement several times over would be feasible without changing the measuring set-up. We showed that the neutron flux through our scatterer/mirror system depends on the horizontal neutron velocity and we measured a small shift in the dependence of the neutron flux through the slit between a mirror and scatterer as a function of the slit size following a change in the horizontal velocity component. The result (0.16±0.04 μm/ms) agrees with the theoretical expectation of 0.2 μm/ms for an average observation time of neutrons in the slit between a mirror and a scatterer equal to $\tau^{pass}$ =15 ms. Methodical errors of the present method, as well as the ways to reduce them, are investigated. We show that the spatial resolution of the method is close to its theoretical limit, defined by the sharpness of the neutron wave functions as a function of height, or, in other words, by the sharpness of the permeability of the gravitational barrier for neutrons as a function of height. The accuracy of this experiment could be improved if a more precise model of the neutron's interaction with a scatterer were developed and the scatterer positioning accuracy improved. A more significant increase in the accuracy of experiments of this kind could be achieved by the long storage of neutrons in quantum states and by a measurement of the frequency of resonant transitions between them, thus allowing one directly to calculate the energies of the corresponding quantum states.

We are grateful to K. Ben-Saidane, F. Descamps, P. Frauenfeld, P. Geltenbort, A. Hillairet, M. Jentschel, M. Klein, T. K. Kuzmina, J. F. Marchand, V. A. Rubakov, R. Rusnyak, M. E. Shaposhnikov, C. Schmidt, I. A. Snigireva, S. M. Soloviev, A. V. Strelkov, P. G. Tinyakov, van der Yver, and v. Walter for their advice and help in the preparation of this experiment and in the measurement. The present work was supported by the INTAS grant 99-705 and by the German Federal Ministry for Research and Education under contract 06HD1531. We are sincerely grateful to those who have shown interest in this work and contributed to its development.

*Figure 1*
Schematic view of the present experiment. From left to the right: the vertical bold lines indicate the upper and lower plates of the input collimator (1); the solid arrows correspond to classical neutron trajectories (2) between the input collimator and the entry slit between a mirror (3, empty rectangle below) and a scatterer (4, black rectangle above). The dotted horizontal arrows illustrate the quantum motion of neutrons above a mirror (5), and the black box represents the neutron detector (6). The size of the slit between the mirror and scatterer could be changed and measured.

*Figure 2*
A typical surface profile of a scatterer with a cooper coating, measured using an atomic-force microscope. The maximum height difference between two points here is equal to 2 μm, the distance between neighbouring "hills" is 7 μm.

*Figure 3*
Schematic view of the integral method for measuring the neutron horizontal velocity component spectrum along the neutron beam axis and for shaping this spectrum. The black vertical lines on the left show the upper plate (1) and the lower plate (2) of the entrance collimator. Each plate could be adjusted to the required height. The jagged black rectangle on the top indicates a scatterer (3); the incomplete grey rectangle bellow shows the mirror (4). The distance between the entry collimator and mirror is equal to $l$. The neutron trajectory rises by $\Delta l$ at the horizontal length $l$. The neutron trajectory is horizontal at the entrance to the mirror/scatterer slit.

*Figure 4*
A measurement of the spectrum of neutron horizontal velocity components. The circles show results of the spectral measurement using the upper collimator plate (the lower plate of the entrance collimator is installed at its lowest position). The stars indicate results of the spectral measurements using the lower collimator plate (the upper plate of the entrance collimator is installed at its highest position). The solid lines approximate the experimental data with Boltzman distribution (for simplicity) in both cases.

*Figure 5*
Set-up for measuring the probability of neutron loss from specular trajectories on their reflection from a mirror. The empty rectangles represent two parallel mirrors (1); the arrows indicate the direction of the neutron beam axis (2); the black box represents a neutron detector (3), placed at a distance of 4 cm from the mirrors and protected by a horizontal diaphragm (4) with a narrow slit of vertical size ±2 mm and of height equal to that of the slit between the mirrors.

*Figure 6*
Estimation of the neutron losses from specular trajectories $K_{loss}(\Delta z)$ using their transmission through a slit between two parallel mirrors at a small grazing angle as a function of the slit size. One mirror is of polished glass, another mirror is of similar glass *coated with copper (circles) or with Ti-Zr-Gd (stars). The horizontal lines correspond to the probabilities measured $K_{loss}(\infty)$.*

*Figure 7*
*The probability of neutron tunneling through the gravitational barrier as a function of height $z$ above a critical value $z_n$.*

*Figure 8*
*Oscillation frequency of an $RC$-generator with capacity for calibrating as a function of the distance between mirror and scatterer in the range relevant to the present experiment with neutrons. The circles show results of calibration using wire-spacers of known thickness. The stars indicate results using a long-focus microscope. The boxes correspond to a measurement using a mechanical comparator. The solid line approximates the data (this line is defined by the most precise measurement with wire-spacers).*

*Figure 9*
*A dependence on slit size of the neutron flux through a slit between a mirror and a copper-coated scatterer. The circles indicate the experimental results obtained with a broad spectrum of horizontal velocity components along the beam axis (Figure 3) with an average value of 6.5 m/s. The stars show an analogous measurement with the soft part of this spectrum with the average value of 4.9 m/s. The solid lines correspond to the classical expectations for these two measurements, normalized in such a way that they correspond to the experimental data at large slit sizes. The horizontal lines indicate the detector background and its uncertainty.*

*Figure 10*
*The dependence of the neutron flux through a slit between a mirror and a copper-coated scatterer as a function of slit size. The rectangles show the results of a measurement with an average horizontal neutron velocity component of 7.8 m/s; the circles – 6.5 m/s; the stars – 4.9 m/s. The solid, dotted and dashed lines correspond to the corresponding simplified quantum-mechanical dependence $F(\Delta z) \sim (\Delta z - X)^{3/2}$. The first and third data sets (corresponding to the collimated neutron beam) are normalized in such a way that the average neutron count rate (over the slit size) would be equal to an analogous value obtained using the second data set (corresponding to the broad neutron spectrum). The horizontal lines indicate the detector background and its uncertainties.*

*Figure 11*
*Dependence of the neutron flux through a slit between a mirror and a scatterer as a function of the slit size. The stars show results of a measurement with a copper-coated rough scatterer. The circles indicate a measurement with Ti-Zr-Gd-coated rough scatterer. The solid line corresponds to the classical expectation normalized so that it approximates the data at large slit sizes. The horizontal lines indicate the detector background and its uncertainty. The average horizontal neutron velocity component along the neutron beam axis is equal to 4.9 m/s.*

*Figure 12*

*Dependence of the neutron flux through a slit between a mirror and a scatterer as a function of the slit size. The circles show the summary results obtained with copper-coated and Ti-Zr-Gd-coated rough scatterers. The solid black curve corresponds to the classical expectation normalized so that it approximates the experimental data at large slit sizes. The dotted line illustrates a simplified quantum-mechanical dependence, which assumes an existence of the lowest quantum state alone and the classical asymptotics at large slit sizes. The horizontal lines indicate the detector background and its uncertainty. The average horizontal neutron velocity component along the neutron beam axis is equal to 4.9 m/s. The solid curve approximates the experimental data with a quantum-mechanical dependence (eq.(22)), in which the heights of the first and second quantum states, the lowest quantum state population, the normalization constant, and the scatterer efficiency are free parameters.*

*Figure 13*

*Schematic view of the experiment with a long bottom mirror (1, shown as the empty box) and two scatterers (2, 3, shown as black boxes). The first scatterer (2, left) shapes the neutron spectrum. It is installed at the constant height of 42 μm. The second scatterer (3, right) analyses the resulting neutron spectrum. Its height is varied. The detector (4), shown as the black box, measures the total neutron flux at the exit of the slit between the mirror and the analyzing scatterer. The distance between the scatterers is equal to 9 cm.*

*Figure 14*

*The circles show results obtained with the long mirror and two scatterers installed as in Figure 13. The neutron flux was measured as a function of the slit size between the mirror and the analyzing scatterer. The solid and two dotted horizontal lines show the neutron count rate at large ("infinitely large") slit size. The solid vertical line indicates the size of the slit between the mirror and the first shaping scatterer. The two vertical dotted lines correspond to the average scatterer roughness amplitude. The inclined solid curve illustrates the neutron count rate in the absence of the first shaping scatterer.*

*Figure 15*

*These results of a measurement of the neutron density above a mirror in the Earth's gravitational field were obtained using a high-resolution plastic nuclear-track detector with uranium coating. The horizontal axis corresponds to a height above a mirror in microns. The vertical axis gives the number of events in an interval of heights. The solid line shows the theoretical expectation under the assumption that the spatial resolution is infinitely high. The populations of the quantum states calculated correspond to those measured by means of two scatterers using the method shown in Figure 13.*

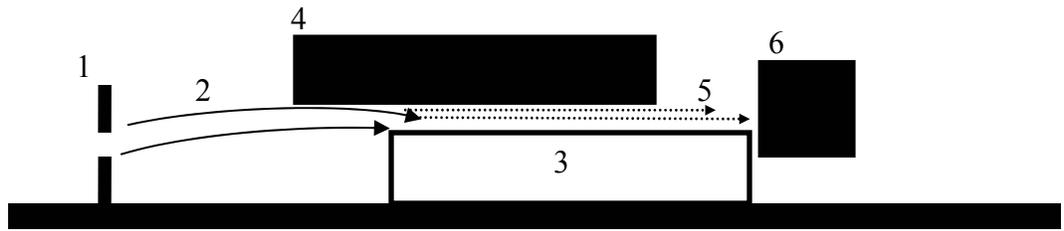

*Figure 1*

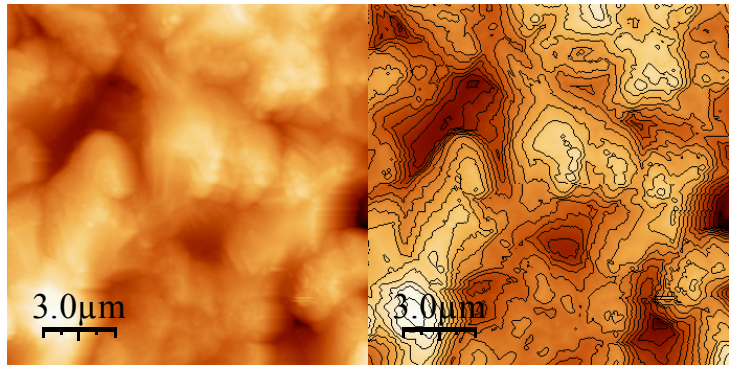

*Figure 2*

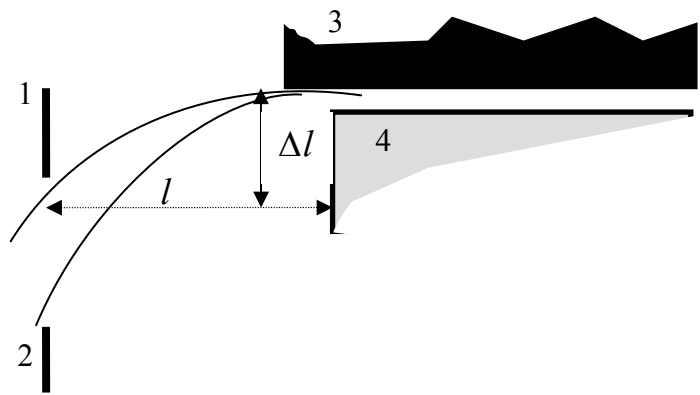

*Figure 3*

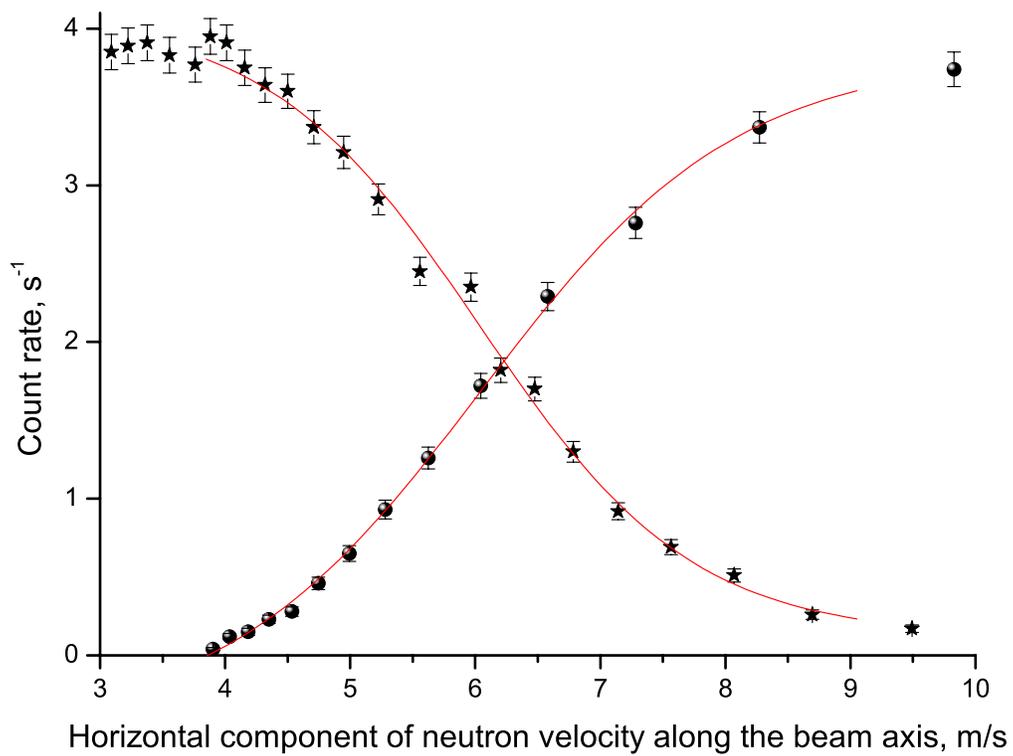

*Figure 4*

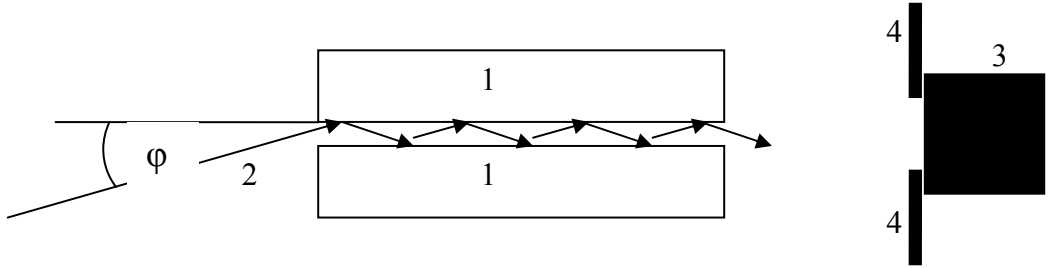

*Figure 5*

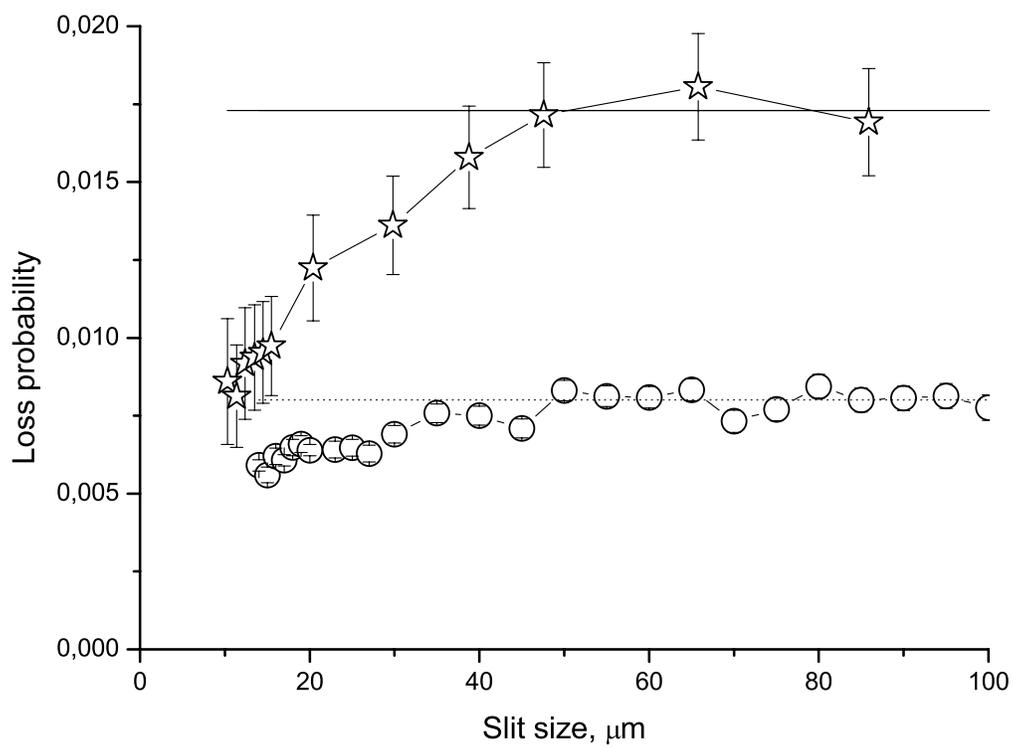

*Figure 6*

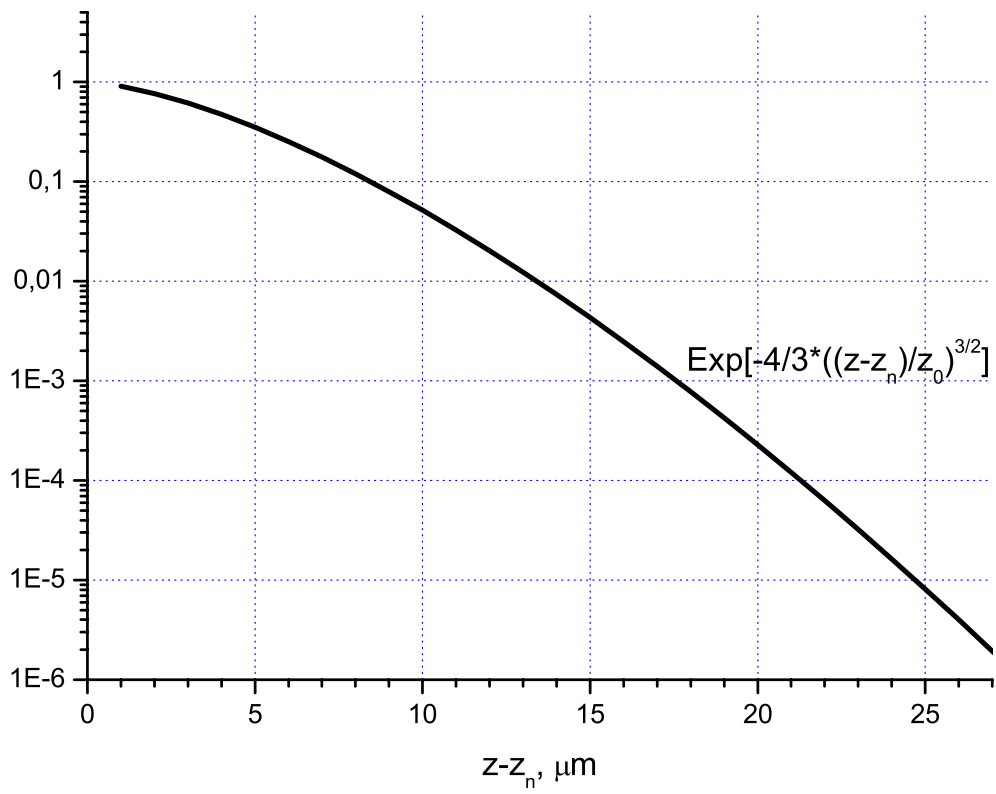

*Figure 7*

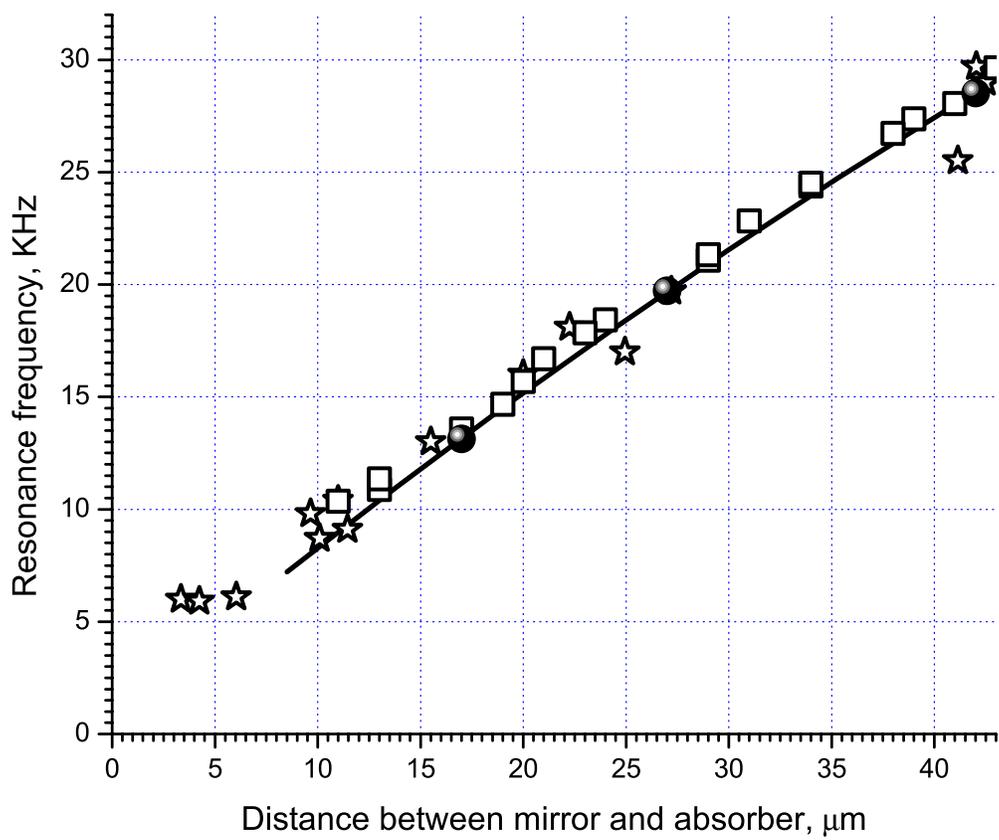

*Figure 8*

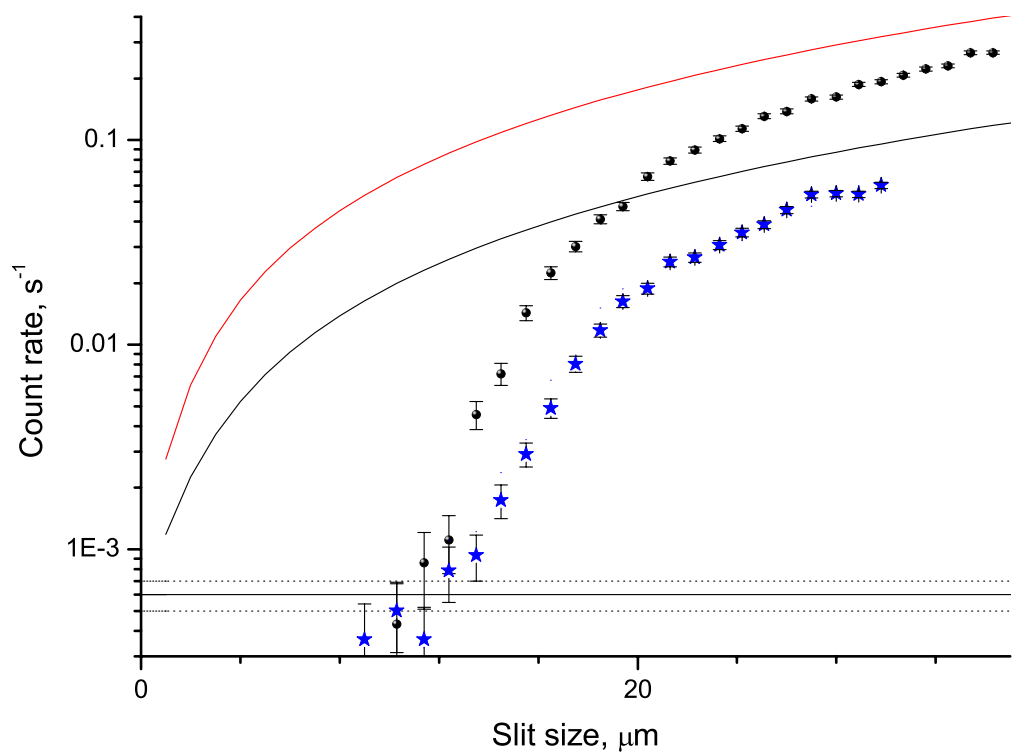

*Figure 9*

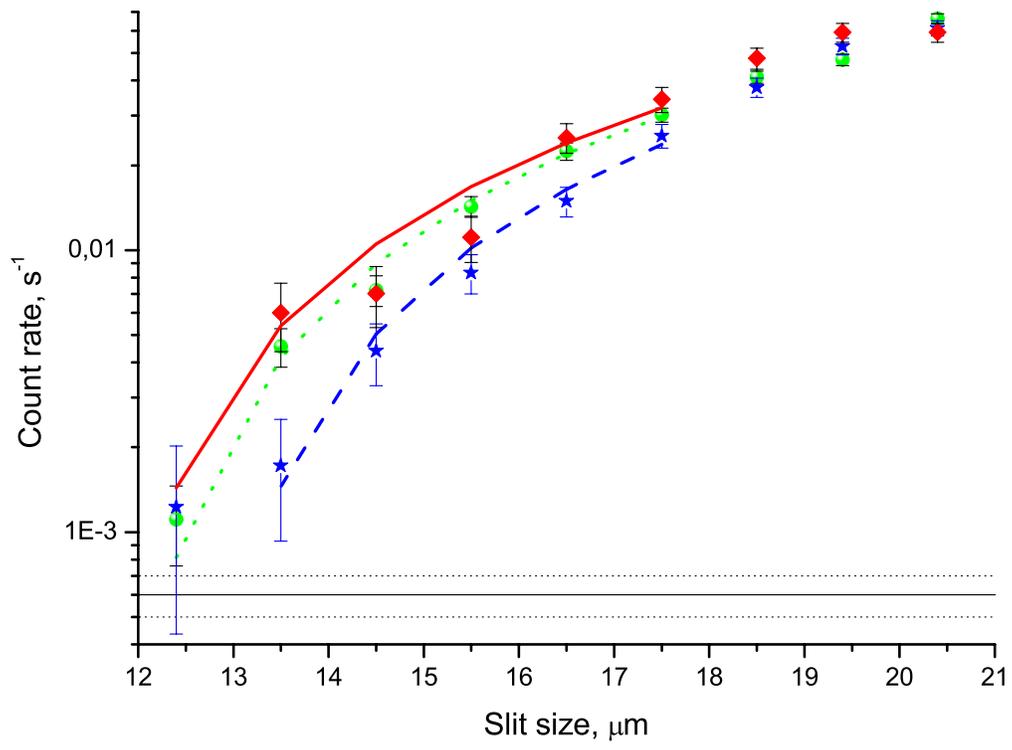

*Figure 10*

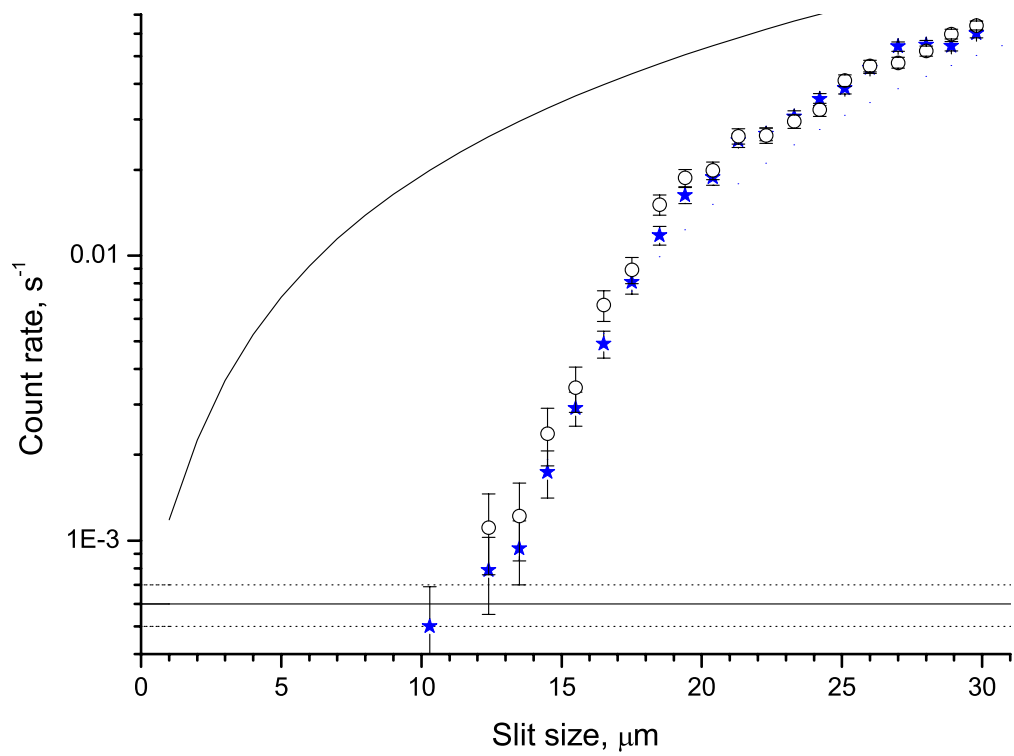

*Figure 11*

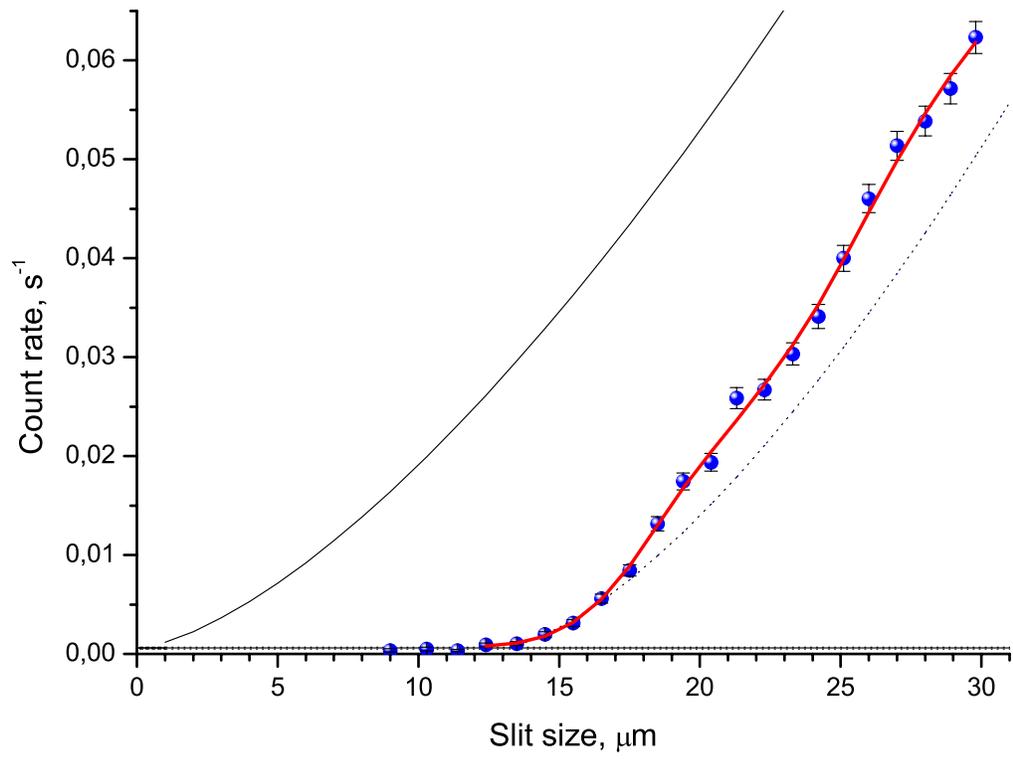

*Figure 12*

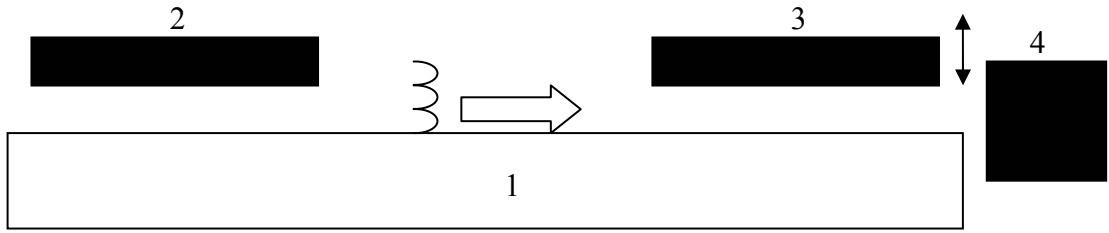

*Figure 13*

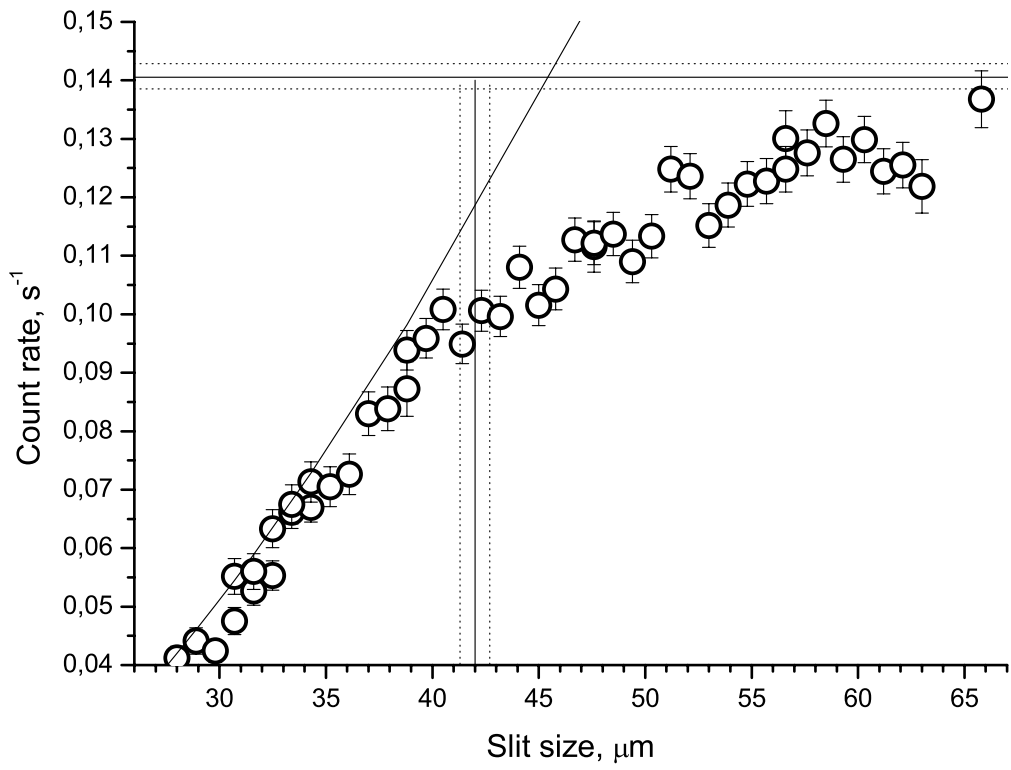

*Figure 14*

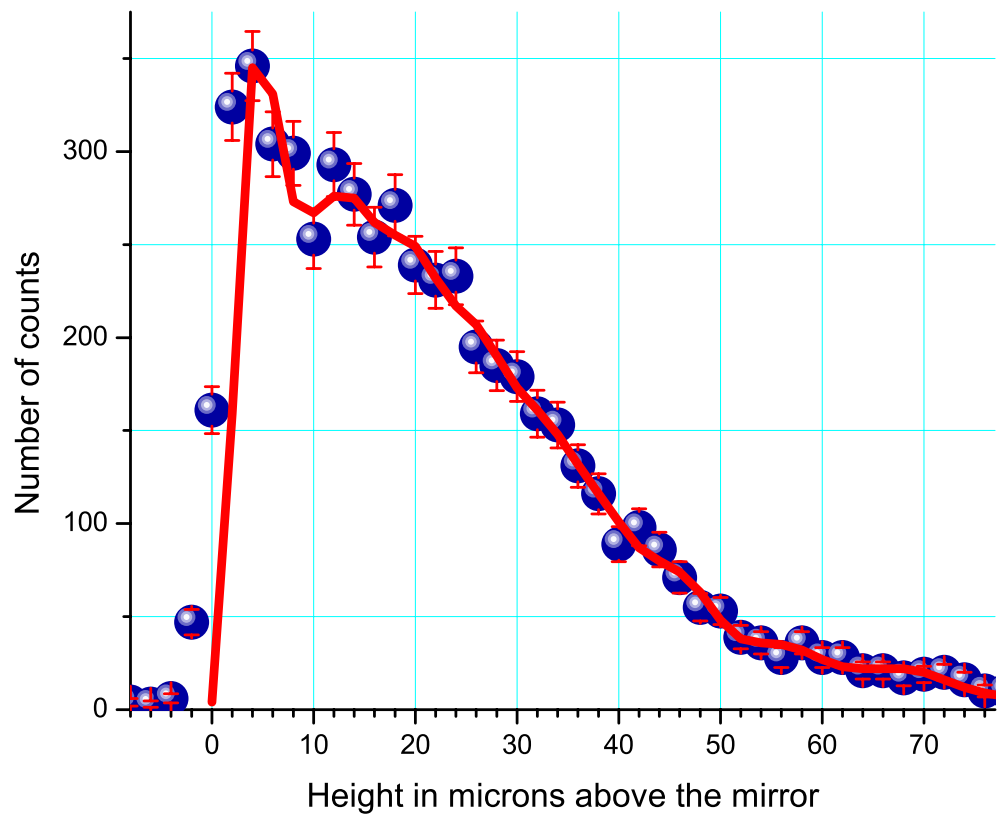

*Figure 15*